\documentclass[11pt,a4paper]{article}
\usepackage[T1]{fontenc}
\usepackage[utf8]{inputenc}
\usepackage[sc]{mathpazo}

\usepackage[a4paper,margin=2.5cm]{geometry}
\usepackage{amsmath,amssymb}
\usepackage{graphicx}
\usepackage{booktabs,tabularx,array}
\usepackage{float}
\usepackage[labelfont=bf,labelsep=period,font=small]{caption}
\usepackage[numbers,sort&compress,square]{natbib}
\usepackage[colorlinks=true,linkcolor=blue,citecolor=blue,urlcolor=blue]{hyperref}

\makeatletter\renewcommand{\p@subsection}{}\makeatother
\newcommand{\cQ}{c_\mathrm{Q}}
\newcommand{\CQ}{C_\mathrm{Q}}
\newcommand{\EF}{E_\mathrm{F}}
\newcommand{\DT}{D_\mathrm{T}}
\newcommand{\TABLEWIDTH}{}
\newcommand{\TABLEFULLWIDTH}{\textwidth}
\newcommand{\FIGWIDTH}{\textwidth}

\title{From Integer Quantized Hall Plateaus to $\mathbb{Z}_2$ Topological Surfaces:\\ Quantum Capacitance as a Probe of the Boundary Density of States}
\author{Afif Siddiki\\[4pt]
\small Vocational School, Atlas University, Anadolu Str.~4, 34480 Istanbul, T\"urkiye\\
\small \href{mailto:afif.siddiki@atlas.edu.tr}{afif.siddiki@atlas.edu.tr}}
\date{}

\begin{document}
\maketitle

\begin{abstract}
The integer quantum Hall effect (IQHE), $\mathbb{Z}_2$ topological insulators (TIs), and quantum capacitance are generally treated as separate subjects, yet share one thing: bulk topology gives rise to a protected boundary state, and quantum capacitance is one of the few electrostatic methods able to detect the density of states (DOS). This paper examines that connection, restricted to single-particle band topology; the fractional QHE is a many-body phase outside this scope, treated separately. The IQHE showed quantized boundary transport has a topological (Chern-number) origin, extended to zero field in $\mathbb{Z}_2$ insulators with helical edge and Dirac surface states. Quantum capacitance has become a key technique for mapping the Landau-level bulk DOS underlying the QHE and the boundary Dirac DOS of topological surfaces. Self-consistent screening theory complicates this: the bulk divides into compressible and incompressible regions shifting with field, and local impedance measurements read as quantum capacitance show the bulk is not uniformly incompressible even while the plateau remains quantized. The paper maintains quantum capacitance is one of the clearest windows into the boundary DOS that topology imposes---while remaining insensitive to the invariant itself---and reveals the local, field-dependent character of the `insulating bulk' assumption behind the QHE--TI analogy.

\end{abstract}

\section{Introduction}\label{sec:intro}

The quantum Hall effect (QHE), topological insulators (TIs), and quantum
capacitance are generally presented in textbooks as three distinct
topics---the quantum Hall effect as a transport phenomenon,
topological insulators as a class of materials, and quantum capacitance
as a device-physics concept. In reality, however, they are closely
related. The QHE was the first experimental example in which a bulk
topological invariant controlled a measurable, quantized physical
quantity~\cite{vonKlitzing1980,Tsui1982,Thouless1982}. Topological insulators have extended this concept
to the case of zero magnetic field by employing spin--orbit coupling in
place of a Landau-quantizing field to generate topologically protected
boundary states~\cite{Kane2005,Bernevig2006,Koenig2007,Zhang2009,XiaY2009,Hasan2010}. Although quantum capacitance is not a
topological quantity by itself, it is one of the very few electrostatic
quantities that are directly sensitive to the density of states (DOS)
resulting from these topological band structures---which makes it a
suitable diagnostic tool for both types of systems.

The claim made here is deliberately bounded. Quantum capacitance is not
a measurement of topology: neither the Chern number nor the $\mathbb{Z}_2$ index can
be extracted from a compressibility measurement. What quantum
capacitance probes is the density of states that a given invariant
enforces at the boundary---and it does so across systems that are
otherwise studied with entirely different techniques. That is the sense
in which it is described here as unifying, and the distinction is
developed in Section~\ref{sec:ti} and revisited in Section~\ref{sec:cannot}.

\section{The Quantum Hall Effect and Its Topological Origin}\label{sec:qhe}

Von Klitzing, Dorda, and Pepper found that the Hall resistance of a
two-dimensional electron gas (2DEG) in a strong perpendicular magnetic
field shows plateaus at the values $h/\nu e^2$, where the integer $\nu$ is the
filling factor, and that these values are independent of the sample's
geometry or degree of disorder~\cite{vonKlitzing1980}. Two years later, Tsui, St\"ormer,
and Gossard observed additional plateaus at fractional filling factors
in higher-mobility samples at lower temperature---the fractional quantum
Hall effect, which arises from a correlated electron liquid rather than
from any splitting of the integer plateaus~\cite{Tsui1982}. The fractional
effect is a many-body phase and lies outside the scope of the present
discussion; the reasons are set out in Section~\ref{sec:cannot}.

The integer quantum Hall plateaus were later understood not as an
accident of a particular material but as a consequence of a topological
invariant---the first Chern number, or TKNN invariant, of the filled
Landau-level bands~\cite{Thouless1982}. Each filled, non-degenerate Landau level
contributes exactly one quantum of Hall conductance---two, where spin
degeneracy is unresolved---and for that number to change, the bulk
energy gap would have to close. The robustness of the plateaus against
disorder is thus a manifestation of topological robustness and not of
the material's purity; Laughlin's gauge
argument had already established this robustness on general grounds,
without reference to a specific band structure~\cite{Laughlin1981}. The insight
that a bulk topological index determines the number of protected,
dissipationless edge channels is the conceptual basis upon which the
field of topological insulators developed, and the transport
consequences of those channels were formalized in the multiprobe
scattering picture of Landauer and B\"uttiker~\cite{Buttiker1988}.

This statement requires one refinement that matters for what follows. In
a disordered sample the Landau levels are broadened into bands in which
only states near the centre are extended, the remainder being localized
by the random potential~\cite{Aoki1981,Prange1981}. The quantized value of
$\sigma_{xy}$ is protected as long as no extended states lie at
the Fermi energy---that is, by a \emph{mobility} gap rather than a
spectral gap. The distinction is not a technicality. Localized states
are what give the plateaus their finite width, since they absorb the
change in carrier density as the field is swept and pin
$\EF$ between Landau-level centres; a disorder-free system
would show quantization only at isolated fields. They also carry
thermodynamic weight: $\partial n/\partial\mu$ does not vanish when $\EF$ lies
among them. Exact quantization of the Hall resistance is therefore
compatible with a bulk of finite thermodynamic density of states, and
the two should not be conflated~\cite{Aoki1981,Prange1981,Halperin1982,Huckestein1995}.

\section{From Landau Levels to Topological Insulators}\label{sec:ti}

The natural question was whether a similar bulk--boundary correspondence
could arise in the absence of an external magnetic field. Kane and Mele
demonstrated theoretically that spin--orbit coupling in systems similar
to graphene could lead to a ``quantum spin Hall'' state: a
time-reversal-symmetric version of the quantum Hall effect in which
electrons with spin up and those with spin down travel in opposite
directions along the edge, the state being protected by a
$\mathbb{Z}_2$ topological invariant rather than a Chern number~\cite{Kane2005}.
Bernevig, Hughes, and Zhang then predicted that the HgTe/CdTe
quantum wells could realize this state~\cite{Bernevig2006}, a prediction which K\"onig
et al.\ confirmed experimentally by means of quantized edge conduction~\cite{Koenig2007}.

The idea soon extended to three dimensions. Zhang et al.\ predicted that
Bi$_2$Se$_3$, Bi$_2$Te$_3$, and Sb$_2$Te$_3$ possess a single gapless Dirac cone in their
spin-momentum-locked surface states which wrap an insulating bulk, while
Y.~Xia et al.\ confirmed this using angle-resolved photoemission
spectroscopy (ARPES)~\cite{Zhang2009,XiaY2009}. The Colloquium by Hasan and Kane, along
with the complementary review by Qi and Zhang, established the general
classification of these phases and their connection to the integer QHE,
specifying the 2D topological insulator as ``a close cousin of the
integer quantum Hall state'' and the 3D case as one that exhibits a
topological magnetoelectric response~\cite{Hasan2010,Qi2011}.

\begin{figure}[H]
\centering
\includegraphics[width=\FIGWIDTH]{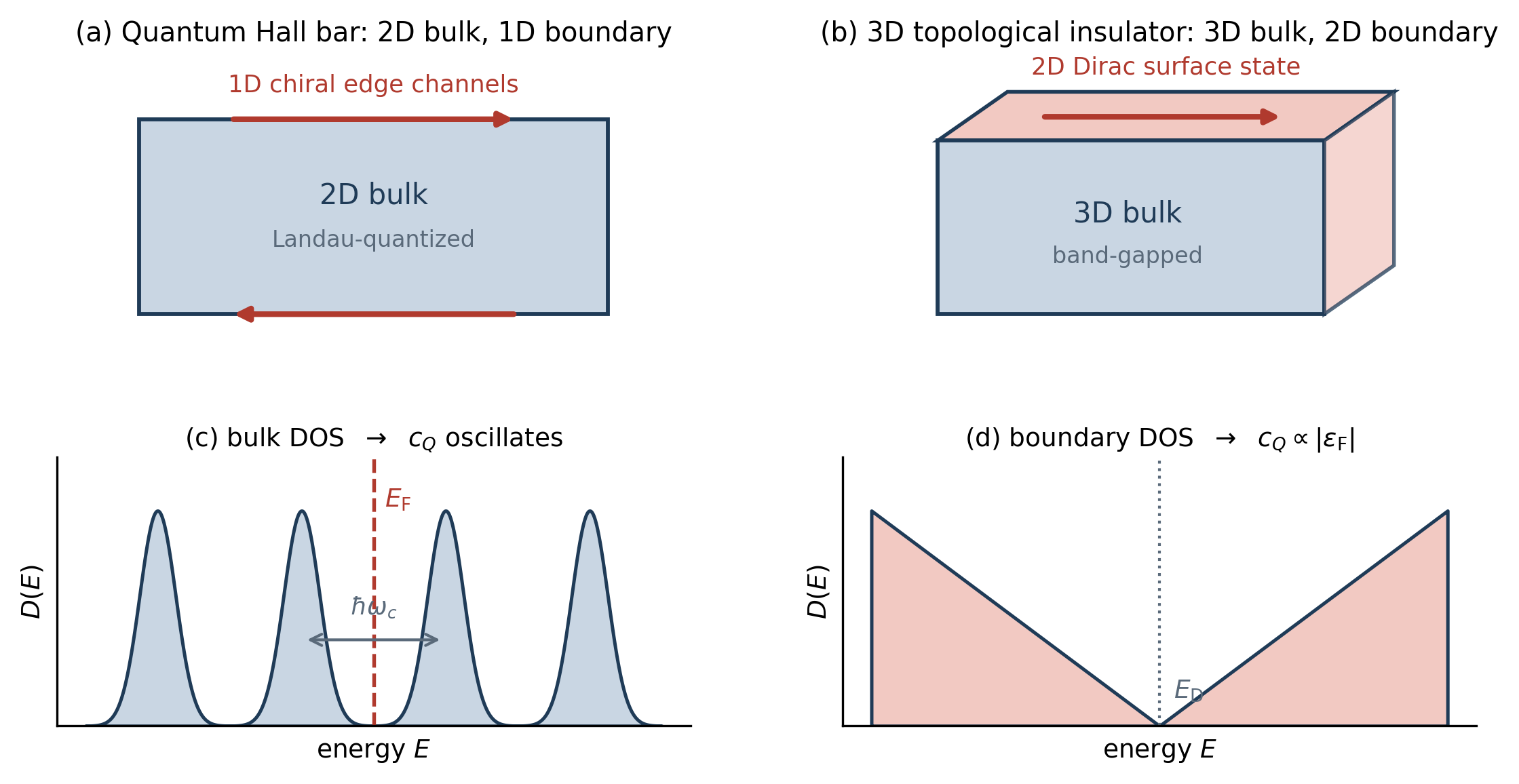}
\caption{The bulk--boundary structure of the two systems and
the density of states a capacitance measurement reaches in each. (\textbf{a}) A
quantum Hall bar is a two-dimensional bulk with a one-dimensional
boundary: the chiral edge channels run along the perimeter and occupy a
negligible fraction of the gated area, so the measurement reads the
bulk. (\textbf{b}) A three-dimensional topological insulator inverts the
arrangement: the bulk is band-gapped and the protected Dirac state
occupies the entire surface, so the measurement reads the boundary.
(\textbf{c}) The corresponding bulk density of states in the quantum Hall case, with
$\EF$ lying in a cyclotron gap, so that $\cQ$
oscillates as the field is swept. (\textbf{d}) The boundary density of states of
the topological insulator, linear in energy about the Dirac point
$E_\mathrm{D}$, giving $\cQ \propto |\varepsilon_\mathrm{F}|$.
Panels (\textbf{a}) and (\textbf{b}) are schematic; no scale is implied.\label{fig:geometry}}
\end{figure}

The main point is that, in the case of the quantum Hall effect, a bulk
Chern number determines the boundary conductance; in topological
insulators, it has been shown that time-reversal-symmetric spin--orbit
coupling can lead to a similar zero-field boundary state governed by a
$\mathbb{Z}_2$ invariant. In both cases one has a gapped bulk and a
metallic boundary. The two systems are not, however, symmetric in what a
capacitance measurement can reach, and it is worth being explicit about
this before the analogy is pushed further. The quantum Hall system is a
two-dimensional bulk with a one-dimensional boundary: the chiral edge
channels occupy a line at the sample perimeter and contribute negligibly
to the capacitance of a gated Hall bar, so a capacitance measurement on
a 2DEG reads the bulk spectrum---the Landau levels---and not the
protected edge states themselves. A three-dimensional topological
insulator inverts this: the bulk is nominally gapped and inert, while
the protected Dirac state occupies the entire gated surface, so the same
measurement reads the boundary. The Landau-level density of states and
the Dirac surface density of states are therefore not two instances of
the same object. The former is the bulk-gap signature of the quantum
Hall system, and the latter the boundary signature of the topological
insulator (Figure~\ref{fig:geometry}).

\begin{figure}[H]
\centering
\includegraphics[width=\FIGWIDTH]{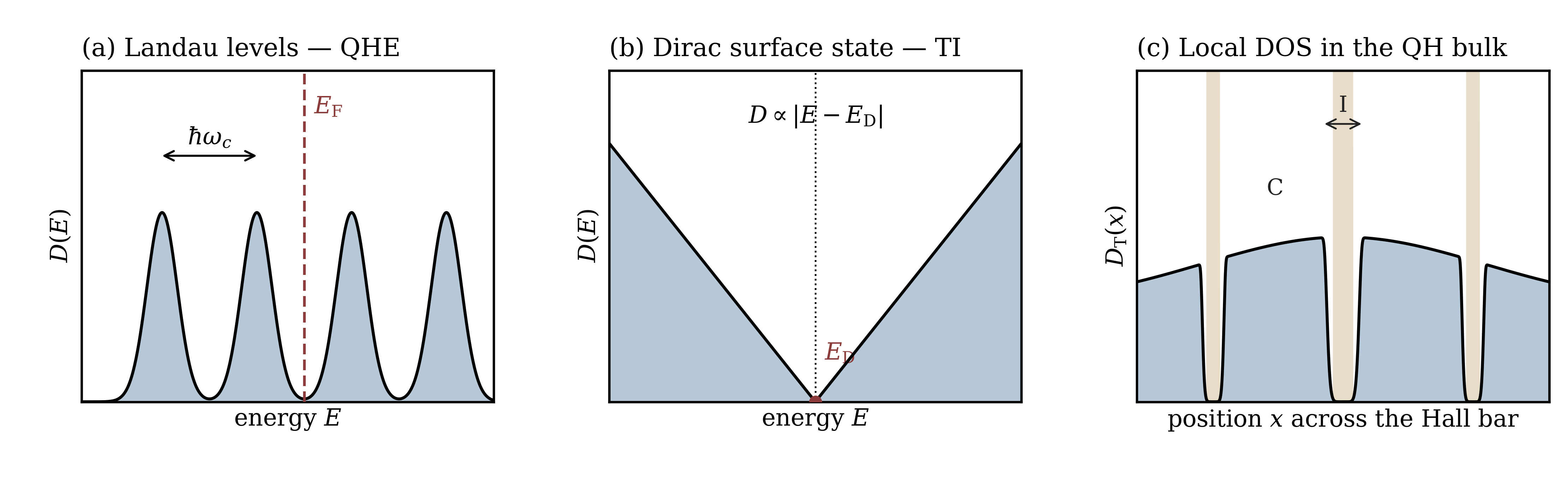}
\caption{The three density-of-states profiles at issue. (\textbf{a}) In
the quantum Hall effect, the bulk DOS collapses into broadened,
macroscopically degenerate Landau levels separated by cyclotron gaps, so
the quantum capacitance oscillates as $\EF$ is swept
through them. (\textbf{b}) The topological surface state of a three-dimensional
TI has a linear Dirac DOS vanishing at $E_\mathrm{D}$, giving
$\cQ \propto |\varepsilon_\mathrm{F}|$. Panels (\textbf{a}) and
(\textbf{b}) are functions of energy and refer to the idealized, translationally
invariant bulk that the topological classification describes. (\textbf{c}) The
same quantity resolved in real space across an actual Hall bar:
self-consistent screening splits the sample into compressible regions
(C) with a finite thermodynamic DOS and incompressible strips (I,
shaded) with $\DT = 0$ and therefore zero quantum
capacitance, whose widths and positions move continuously with magnetic
field while the global topological index and the quantized plateau stay
fixed. Axes are schematic; no scale is implied.\label{fig:dos}}
\end{figure}

What is common is the role played by the density of states rather than
the location at which it is measured. In each system the topology fixes
a gapped bulk and a protected boundary, and quantum capacitance reads
out whichever member of that pair is electrostatically accessible in the
given geometry. In the quantum Hall case, that is the bulk: the
vanishing density of states when $\EF$ lies in a cyclotron
gap is precisely the condition under which the Chern number is well
defined and the plateau is quantized, so the Landau-level oscillation is
a thermodynamic verification of the gap that guarantees the edge
channels, observed in place of the channels themselves. In the
topological insulator case it is the boundary: the linear Dirac density
of states is the state that the bulk $\mathbb{Z}_2$ index
guarantees, measured directly. Figure~\ref{fig:geometry} and Table~\ref{tab:signatures} set out the
correspondence; Figure~\ref{fig:dos} shows the three density-of-states profiles that
follow from it. The one case in which this geometric restriction has
been partly overcome is discussed in Section~\ref{sec:local}, where contacts placed
inside the Hall bar give the capacitance measurement spatial resolution
within the quantum Hall bulk itself.

\begin{table}[H]
\caption{Bulk and boundary density-of-states signatures in the
two systems, and which of them a quantum-capacitance measurement reaches
in each case. The asymmetry in the last row is geometric rather than
conceptual: the quantum Hall boundary is one-dimensional and occupies a
negligible fraction of the gated area, whereas the topological-insulator
boundary is two-dimensional and occupies all of it.\label{tab:signatures}}
\TABLEWIDTH
\small
\begin{tabularx}{\TABLEFULLWIDTH}{>{\raggedright\arraybackslash}p{0.2\TABLEFULLWIDTH}
  >{\raggedright\arraybackslash}X >{\raggedright\arraybackslash}X}
\toprule
 & \textbf{Quantum Hall effect (2DEG)} & \textbf{Three-dimensional topological insulator} \\
\midrule
Invariant & Chern number (TKNN) & $\mathbb{Z}_2$ index \\
\midrule
Bulk & Two-dimensional; Landau-quantized, gapped when $\EF$
lies in a cyclotron gap & Three-dimensional; band-gapped \\
\midrule
Bulk DOS signature & Landau-level oscillations; DOS suppressed but not
zero within the mobility gap (localized states) & DOS $\to 0$ within the
band gap (ideal case) \\
\midrule
Boundary & One-dimensional chiral edge channels & Two-dimensional Dirac
surface state \\
\midrule
Boundary DOS signature & Edge-channel DOS; negligible gated area &
Linear Dirac DOS, $\cQ \propto |\varepsilon_\mathrm{F}|$ \\
\midrule
What $\cQ$ measures in practice & The bulk: Landau levels~\cite{Smith1985};
locally resolved within the bulk~\cite{Kendirlik2017} & The boundary:
Dirac surface state~\cite{Inhofer2018,Xiu2012} \\
\bottomrule
\end{tabularx}
\end{table}

It is worth being specific here about what quantum capacitance, the
thermodynamic quantity $\cQ = e^2(\partial n/\partial\mu)$ formally
introduced in Section~\ref{sec:qc}, does and does not do. A topological invariant,
like the Chern number for the quantum Hall effect or the
$\mathbb{Z}_2$ index for a topological insulator, is an integer that
is a globally defined property of the filled bulk band structure. It is
quantized, cannot change by smooth deformations of the Hamiltonian, and
can only change if a bulk gap closes. It is not itself an observable
quantity in the thermodynamic sense: the Chern number is not directly
measured, but rather deduced from the number and chirality of the
protected boundary channels it guarantees. Quantum capacitance, in
contrast, continuously probes the density of states at the Fermi level
as a function of energy. In a bulk gap, the presence of the invariant is
reflected in the density of states only indirectly, through the
existence and dispersion of the boundary states that the gap protects,
and through the vanishing (or finite) density of states in the bulk
itself. Thus, the measured density of states profile is a reflection of
the underlying topology, not the invariant itself, and one should not
treat the two as if the quantum capacitance were ``measuring'' the
topological number.

\section{Quantum Capacitance as a Probe of the Density of States}\label{sec:qc}

Luryi explained that a two-dimensional electron system does not exhibit
the behavior of an ideal metallic capacitor plate since its density of
states is finite and therefore the applied field only partially screens
the system. This adds a further (quantum) capacitance in series with the
geometric one,
\begin{equation}
\cQ = e^2\,\frac{\partial n}{\partial \mu},
\label{eq:cq}
\end{equation}
which is proportional to the density of states at the Fermi level ($\EF$)~\cite{Luryi1988}.
Here $n$ is the areal carrier density, so that $\cQ$ is a
capacitance per unit area; the quantum capacitance of a region of area $A$
is $\CQ = \cQ A$. Lower-case $\cQ$ is used throughout for the intensive quantity and
upper-case $\CQ$ where the capacitance of a specific region
or device is meant. Unlike transport measurements, which depend on
scattering and mobility, quantum capacitance is a thermodynamic,
near-equilibrium measurement: it directly indicates compressibility and
the density of states, even in samples so disordered that clean
transport cannot take place.

In graphene and in other two-dimensional electron gases, this method
shows the density of states with great clarity---for example, J.~Xia et
al.\ employed quantum capacitance to map the density of states of
graphene, resolving both its linear dependence on energy and its finite
residual value at the Dirac point~\cite{XiaJ2009}. When a magnetic field is
applied, the quantum capacitance oscillates in accordance with the
discrete density of states of the Landau levels, offering an independent
thermodynamic check of the Landau quantization responsible for the
quantum Hall effect and complementing measurements of Hall-plateau
transport~\cite{Smith1985}.

\section{Quantum Capacitance of Topological Surface States}\label{sec:tss}

The same reasoning holds true for TI surface states. Since the
topological surface state has a linear Dirac dispersion, its density of
states is zero at the Dirac point and increases linearly with
$|E - E_\mathrm{D}|$, which results in $\cQ \propto |\varepsilon_\mathrm{F}|$,
as opposed to the constant density
of states (and therefore constant $\cQ$) found in a
conventional 2DEG. Experiments carried out on Bi$_2$Se$_3$ have made use of
precisely this feature: the RF quantum capacitance measurements allowed
the Dirac velocity of the topological surface state to be obtained
directly from the slope of $\cQ$ with respect to gate
voltage by means of the Berglund integral---the standard
capacitance--voltage inversion of metal--oxide--semiconductor
analysis---without requiring any previous knowledge of the band
structure~\cite{Inhofer2018}. Other capacitance studies have used the
excitation-frequency dependence of $\cQ$ to distinguish the
surface response from that of the bulk in thin-film Bi$_2$Se$_3$ capacitors~\cite{Xiu2012},
while theoretical work on ultrathin films has traced how
Zeeman coupling and inter-surface hybridization govern the crossover
between coupled and decoupled top and bottom Dirac surfaces~\cite{Tahir2013}.

The relation $\cQ \propto |\varepsilon_\mathrm{F}|$
is an idealization, and real devices depart from it in several ways that
matter when it is used quantitatively. Charge inhomogeneity broadens the
Dirac point into a distribution of local potentials, so the density of
states does not reach zero and the measured capacitance saturates at a
finite minimum rather than closing to a cusp; the same effect is
familiar in graphene, where a finite residual capacitance at the Dirac
point is routinely observed~\cite{XiaJ2009}. Finite temperature acts similarly,
since the measurement averages the density of states over a window of
order $k_\mathrm{B}T$ about the Fermi level, and the depth of the
minimum is set by whichever of disorder broadening and thermal
broadening is larger. Residual bulk carriers are a further complication:
as-grown Bi$_2$Se$_3$ is typically n-type, and bulk states then contribute a
capacitance in parallel with the surface term, so that reaching the
bulk-depleted regime, or separating the two by their different response
times, is a precondition for reading the slope as a surface property~\cite{Inhofer2018,Xiu2012}.
Both surfaces of a slab contribute in parallel, and in thin
films they hybridize, which modifies the spectrum further~\cite{Tahir2013}.
Finally, gating does not shift the Fermi level rigidly. Band bending
near the gated surface can populate trivial two-dimensional
accumulation-layer subbands whose density of states adds to the Dirac
term, and the conversion from gate voltage to $\EF$ through
the Berglund integral is only as reliable as the assumption that the
measured capacitance belongs to the surface state alone. None of this
invalidates the method, but it does mean that an approximately linear
$\cQ$ is a necessary rather than a sufficient indication of
a Dirac surface state, a point returned to in Section~\ref{sec:cannot}.

In the presence of an applied magnetic field, the TI surface states also
give rise to Landau levels, and their quantum capacitance exhibits the
same oscillatory pattern as that observed in ordinary 2DEGs and in
graphene, although the oscillations are predicted to be asymmetrically
centered around the Dirac point and to vary with the square root of the
field rather than with the field itself, which would be a direct
thermodynamic signature of the relativistic Dirac Landau-level spectrum~\cite{Tahir2013}.
This completes the experimental parallel---though at the
level of the density of states, not of the invariant: quantum
capacitance, which was first developed for conventional 2DEGs~\cite{Luryi1988}
and employed to verify the Landau-level density of states associated
with the QHE, is now a standard electrostatic method for characterizing
the Dirac surface states that make topological insulators the zero-field
analogs of the QHE~\cite{Hasan2010}. What such a measurement establishes,
however, is the form of the density of states and not the topology. A
linear, Dirac-like density of states is consistent with a topologically
protected surface state, but it does not by itself fix the $\mathbb{Z}_2$ invariant,
and an unprotected Dirac cone would produce the same signature; the
point is taken up again in Section~\ref{sec:cannot}.

\section{Local Compressibility and the Limits of the Bulk-Insulator Picture}\label{sec:local}

Before turning to the electrostatic picture, it is worth distinguishing
the sense of ``bulk'' used in Section~\ref{sec:ti} from the one used below. The
topological classification is a statement about the global bulk: it
concerns the filled bands of an idealized, translationally invariant
system, and the invariant it defines takes a single value for the sample
as a whole. Local incompressibility, by contrast, is a statement about
the real, finite, electrostatically inhomogeneous sample: it describes
how the density of states and screening response vary from point to
point across the Hall bar, and it can change continuously with magnetic
field and position even while the global topological index stays fixed.
A locally compressible region within the bulk is therefore not a
violation, or even a complication, of the bulk topological gap in the
idealized sense---it reflects the fact that a real device is not
translationally invariant and must be treated electrostatically. The two
descriptions operate at different levels: one classifies the system as a
whole, the other maps its local thermodynamic response, and quantum
capacitance is sensitive only to the latter.

One further distinction should be made explicit, since the terminology
invites conflation. Compressibility, $\partial n/\partial\mu$, is a thermodynamic quantity:
it measures how the electron density responds to a change in chemical
potential. Conductivity is a transport quantity: it measures how current
responds to an applied field. A region of finite $\partial n/\partial\mu$ need not conduct---disorder-localized
states are compressible and insulating at the
same time---and a region of vanishing $\partial n/\partial\mu$ is not thereby shown to
block current along every path. The words \emph{compressible} and
\emph{incompressible} are used in this section in the thermodynamic
sense only, and carry no implication about dc transport. Quantum
capacitance measures the former and says nothing directly about the
latter.

The way Section~\ref{sec:ti} is presented---in terms of a gapped, incompressible
bulk with protected conducting boundaries---is based on a
single-particle approximation. Building on the electrostatic picture of
edge channels introduced by Chklovskii, Shklovskii and Glazman~\cite{Chklovskii1992},
Siddiki and Gerhardts devised a self-consistent screening theory of the
integer quantum Hall effect that treats the two-dimensional electron
system electrostatically at finite temperature; by solving the
Thomas--Fermi--Poisson equation for the density profile it is found that a
Hall bar breaks down into locally compressible regions (which have a
finite thermodynamic density of states) and locally incompressible
strips (which have vanishing density of states and are pinned to an
integer Landau-level filling) whose width and position change
continuously as the magnetic field changes~\cite{Siddiki2004}. In this model the
plateau is sustained by a sufficiently wide incompressible strip
suppressing backscattering between the edges of the sample, not because
the whole bulk is incompressible during the plateau; the extremely
narrow strips predicted by the simple Thomas--Fermi approximation are
spread out when the finite spatial extent of the electron wavefunctions
and the effects of nonlocal transport are taken into account~\cite{Siddiki2004}.

It is worth stating plainly how this description relates to the
conventional one, since the two are sometimes presented as competitors.
They are better understood as answering different questions. The
topological argument of Section~\ref{sec:qhe} explains why the plateau value is
exactly $h/\nu e^2$ and why it is insensitive to sample details; it is
formulated for an idealized, translationally invariant, non-interacting
system. The disorder--localization picture explains why the plateaus
have finite width in magnetic field, since localized states absorb the
change in carrier density and pin $\EF$ between
Landau-level centres~\cite{Aoki1981,Prange1981,Halperin1982,Huckestein1995}; it is a single-particle account in
which disorder is the essential ingredient. The self-consistent
screening picture addresses a third question---where the current flows
and how the density of states is distributed in space---and it is an
interacting, electrostatic, finite-temperature account in which the
confining potential and the Coulomb interaction are essential. None of
these supersedes the others, and a real sample contains both localized
states and a self-consistently screened density profile. Their relative
importance is a matter of sample quality: disorder-dominated,
low-mobility samples are the natural domain of the localization
description, whereas in high-mobility samples, where comparatively few
localized states are available to pin the Fermi energy across a wide
plateau, the electrostatic structure becomes the more informative
description. The measurements discussed below were made on high-mobility
GaAs/AlGaAs material, which is why the screening picture is emphasized
here; this is a statement about the regime under discussion and not a
claim that the localization picture is incorrect.

A finite bulk thermodynamic density of states across a plateau therefore
admits more than one origin, and it is worth separating them.
Disorder-localized states contribute $\partial n/\partial\mu$ throughout the mobility gap,
as described in Section~\ref{sec:qhe}. Self-consistent screening contributes a
spatially structured density of states, with compressible regions of
finite $\partial n/\partial\mu$ separated by incompressible strips whose width and position
track the magnetic field. Thermal broadening contributes a third. These
are not alternatives, and no capacitance measurement separates them by
itself: quantum capacitance responds to $\partial n/\partial\mu$ and is indifferent to
whether the states it counts are localized or extended. What
distinguishes the screening contribution is not its magnitude but its
structure in field and position. Localization produces a smooth
background that varies slowly across a plateau, whereas the
compressible--incompressible pattern produces a signal keyed to where
the incompressible strip lies relative to the contacts, and therefore a
non-monotonic dependence within a single plateau.

This picture provides a prediction that can be directly tested, making
use of the same physical quantity introduced in Section~\ref{sec:qc}: since
incompressible regions have a zero thermodynamic density of states, they
should also have zero quantum capacitance,
\begin{equation}
\CQ = \cQ A = e^2 D_\mathrm{T}(\EF)\,A ,
\label{eq:CQ}
\end{equation}
where $D_\mathrm{T}(\EF) = \partial n/\partial\mu$ is the thermodynamic
density of states at the Fermi energy---the same quantity introduced
in Section~\ref{sec:qc}, Equation~(\ref{eq:cq})---and $A$ is the relevant area. Local probes had already
shown that the quantum Hall bulk is not homogeneous:
single-electron-transistor measurements of the local compressibility
resolved the coexistence of compressible and incompressible regions and
the way localized states rearrange across a plateau~\cite{Ilani2004}. Kendirlik
et al.\ then probed this directly inside the bulk, inserting two further
voltage probes into a high-mobility GaAs/AlGaAs Hall bar and measuring
the local AC impedance between them while the outer contacts
simultaneously recorded the usual, fully quantized Hall plateau~\cite{Kendirlik2017}.
The quantity actually measured is a complex impedance at
finite excitation frequency, not a capacitance read off a gated
parallel-plate structure. Quantum capacitance enters at the level of
interpretation: the reactive part of the inner-contact response is
attributed to the thermodynamic density of states of the region between
the probes, on the argument that an incompressible region with
$D_\mathrm{T}(\EF) = 0$ would present zero
capacitance and hence a diverging impedance. If the bulk were uniformly
incompressible, as is assumed in the simplest single-particle
bulk-topological argument, the inner impedance should therefore have
diverged throughout the entire plateau. Instead, the measured inner
potential difference---and therefore the local quantum capacitance
inferred from it---remained finite throughout, peaking only in a narrow
subsection of each plateau ($\nu = 2$ and $1$, and also at $\nu = 2/3$, for which
see the caveat in Section~\ref{sec:cannot}) and falling away elsewhere within the same
plateau, thus clearly showing that the bulk is compressible over a large
part of the field range in which the Hall resistance is already
perfectly quantized~\cite{Kendirlik2017}. And when the temperature was raised, the
maximum value decreased in proportion to the inverse of the
thermodynamic density of states, which is what the quantum-capacitance
interpretation predicts and an unrelated cross-coupling would not; this
supports that interpretation, but the density of states remains an
inference from the impedance data rather than a directly recorded
capacitance. This does not exclude a localized-state contribution to the
measured density of states, and none is claimed to be excluded: the
feature relevant here is the variation across the plateau rather than
the finite value itself.

The outcome is in agreement with, not a refutation of, the QHE--TI
analogy presented in Section~\ref{sec:ti}: the Hall plateau remains due to the
presence of incompressible strips and therefore arises from a nonzero
energy gap along some path that either spans or goes around the sample,
this being consistent with a nonzero Chern number~\cite{Siddiki2004,Kendirlik2017}. The local
measurement complicates the assumption of an everywhere incompressible
bulk which underlies the most straightforward formulations of the
bulk--boundary correspondence: in actual devices the boundary between
``protected edge and insulating bulk''~\cite{Buttiker1988} is not sharp but is
instead determined self-consistently by electrostatics, screening, and
the local filling factor. This provides a second, independent example of
the diagnostic capability of quantum capacitance beyond the
topological-insulator surface states discussed in Section~\ref{sec:tss}: just as the
same $\cQ = e^2(\partial n/\partial\mu)$ relationship shows the presence of a
Dirac surface DOS in a TI, it also shows that the ``bulk-insulator''
part of the structural analogy between the QHE and the TI is a local and
field-dependent statement rather than a literal one~\cite{Siddiki2004,Kendirlik2017}.

Independent support for this picture has since come from systems in
which the same question can be asked at zero magnetic field. In
magnetically doped topological insulators exhibiting the quantum
anomalous Hall effect, potentials measured at multiple points within a
device are well described by assuming current flow through the
two-dimensional bulk rather than along edge modes~\cite{Rosen2022}, and
scanning-SQUID magnetic imaging of the current density has directly
visualized a regime in which the sample carries current primarily in its
interior~\cite{Ferguson2023}. The imaging work concludes further that the current
flows through incompressible regions whose spatial structure changes as
the plateau is traversed, and that the self-consistent electrostatics of
the sample play a central role in setting the current distribution.
These are magnetometry and transport experiments rather than capacitance
measurements, and they address the zero-field rather than the
field-induced case, so they do not test the argument made here directly.
They do indicate, however, that the departure from a uniformly
incompressible, edge-only picture is not peculiar to the high-mobility
GaAs system discussed above, and that it appears on the
topological-insulator side of the analogy as well.

\section{What Quantum Capacitance Cannot Do}\label{sec:cannot}

The most fundamental limitation is the one of principle set out in
Section~\ref{sec:ti}: quantum capacitance returns a thermodynamic quantity
evaluated locally and at the Fermi level, not a topological invariant.
It is worth adding what follows from this. The density-of-states
features discussed above---the discrete Landau-level spectrum, the
linear Dirac DOS of a protected surface state, the vanishing DOS of an
incompressible region---are necessary signatures of the corresponding
topological phases, but they are not sufficient ones. A linear density
of states is equally consistent with an unprotected Dirac cone, and a
locally vanishing density of states with trivial localization.
Establishing that a boundary state is topologically protected still
requires spin-resolved photoemission, quantized transport, or
band-structure classification; quantum capacitance is complementary to
those methods rather than a replacement for them.

Advocating quantum capacitance as a probe requires being explicit about
where it is weak. The measured device capacitance is the geometric
capacitance in series with the quantum capacitance, so extracting a
density of states demands independent knowledge of the gate geometry and
the dielectric constant; the two contributions are hardest to separate
precisely where the quantum capacitance is small---near the Dirac point,
and inside an incompressible region---which are exactly the regimes of
interest here. Parasitic capacitance, series resistance and charge
trapping at interfaces distort the measured response at high frequency,
and the finite charging time of a poorly conducting sample can mimic a
vanishing density of states. The measurement also averages over the
electrode area, so a local statement about compressibility is only as
local as the contact geometry permits: the inner-contact arrangement of
Ref.~\cite{Kendirlik2017} pushes that spatial resolution into the bulk, but it is
not a microscope, and there the density of states is extracted from an
impedance rather than read from a gated capacitor, so the frequency and
series-resistance caveats above bear directly on its interpretation.
Quantum capacitance is also blind to localization. It counts states at
the Fermi level without regard to whether they conduct, so a finite
$\cQ$ does not by itself imply a conducting bulk, and a
transport measurement showing $\sigma_{xx} \to 0$ is not in tension
with a capacitance measurement showing $\partial n/\partial\mu \neq 0$ at the same field.
Statements about compressibility and statements about conduction are
separate, and this probe speaks only to the former. None of this
undermines the argument above. It sets out the conditions under which
the measured capacitance can be read as a density of states rather than
as a device response.

\subsection{Scope: Single-Particle Band Topology and the Fractional Regime}\label{sec:scope}

A final limitation is one of scope, and it is worth stating explicitly.
The topological classification used throughout this paper is that of
single-particle band topology: the Chern number of filled Landau-level
bands in the integer quantum Hall effect, and the $\mathbb{Z}_2$
index of a time-reversal-symmetric band insulator. Both are properties
of the occupied bands of an effectively non-interacting system, and both
are what the density-of-states signatures discussed here reflect. The
fractional quantum Hall effect is not of this kind. It is an
intrinsically many-body phase whose ground state is a correlated liquid,
whose gap is a many-body gap rather than a cyclotron or band gap, and
whose elementary excitations carry fractional charge and obey
non-trivial exchange statistics~\cite{Laughlin1983}. It cannot be reached from the
statement that each filled Landau level contributes one conductance
quantum, and the arguments of Sections~\ref{sec:qhe} and~\ref{sec:ti} should not be read as
covering it.

Several frameworks describe the fractional regime, and they differ in
construction rather than in their predictions for the observed fillings.
The Laughlin wave function is a variational ansatz for the many-body
ground state, built from lowest-Landau-level orbitals and Jastrow-type
correlations~\cite{Laughlin1983}. The composite-fermion construction instead
attaches an even number of flux quanta to each electron and maps the
fractional problem onto an effective integer quantum Hall effect of
composite fermions~\cite{Jain2007}. Chern--Simons field theory recasts the same
flux attachment as an effective gauge theory at the level of an action
rather than a wavefunction. The homotopy-based, cyclotron-braid
alternative starts from a different primitive still: it restricts the
exchange trajectories accessible to the particles in the
magnetic-field-constrained multiparticle configuration space, selects
the braid subgroups compatible with that restriction, and recovers
Laughlin--Jastrow correlations and a wider hierarchy from the selected
subgroups~\cite{Jacak2022,Jacak2023}. The four constructions converge closely on which
fillings are allowed and on the existence of a many-body gap; they
differ in what is taken as fundamental---a variational wavefunction, a
mean-field flux attachment, an effective action, or a restricted braid
group---and a systematic comparison of what each predicts beyond the
filling sequence itself (edge structure, non-Abelian generalizations,
and so on) is beyond the scope of the present Perspective. We note these
approaches without adjudicating between them, since none of them is what
quantum capacitance is being used to probe here.

It is worth placing these constructions within a single classification,
since the field-driven and zero-field cases are sometimes discussed as
if the pairing were only IQHE--TI. Single-particle band topology in fact
comes in two field/zero-field pairs. One pair carries a Chern number and
requires broken time-reversal symmetry: the IQHE (Section~\ref{sec:qhe}), realized
with an external magnetic field, and the quantum anomalous Hall effect
or Chern insulator, its zero-field counterpart realized through magnetic
doping or intrinsic magnetism and already encountered in Section~\ref{sec:local}~\cite{Rosen2022,Ferguson2023}.
The other pair carries a $\mathbb{Z}_2$ index and
requires time-reversal symmetry, which forces the Chern number itself to
vanish: this is the topological-insulator family of Section~\ref{sec:ti}, which has
no field-driven analogue for the same symmetry reason. The many-body
counterparts follow the same field/zero-field split. The FQHE (Section~\ref{sec:qhe}
and above in this section) is the field-driven, many-body phase;
fractional Chern insulators (FCIs) are its zero-field, lattice analogue---a
partially filled, topologically nontrivial (Chern-number-carrying)
band stabilized by strong interactions rather than by an external field.
FCIs share the correlated, many-body character of the FQHE rather than
the single-particle band topology on which Sections~\ref{sec:qhe} and~\ref{sec:ti} rest, so the
scope restriction stated above for the FQHE applies to them for the same
reason. What is worth noting is that FCIs have been established in
exactly the thermodynamic quantity this paper is about: local electronic
compressibility measurements in magic-angle twisted bilayer graphene
have resolved a sequence of incompressible states at fractional filling
of a Chern band, each associated with a many-body gap analogous to the
FQHE gap~\cite{Xie2021}. As with the $\nu = 1/3$ GaAs measurement discussed above,
such a measurement establishes the magnitude of a compressibility gap,
not the Chern number of the parent band or the statistics of the
excitations it hosts; the distinction between a thermodynamic signature
and a topological invariant that organizes this entire paper therefore
extends, essentially unchanged, to the fractional Chern regime.

This bears directly on one point in Section~\ref{sec:local}. The inner-contact
measurements were reported at $\nu = 2$ and $\nu = 1$ and also at $\nu = 2/3$, and
the self-consistent screening analysis applied to them is a
single-particle Thomas--Fermi treatment. Extending that analysis to a
fractional filling is not automatic: it requires the composite-fermion
mapping, under which the fractional state of electrons becomes an
integer state of composite fermions and an analogous
compressible--incompressible structure can be defined~\cite{Jain2007}. We
therefore restrict the interpretation offered in Section~\ref{sec:local} to the
integer fillings, and record the $\nu = 2/3$ observation without claiming
that the strip picture as presented accounts for it.

It should not be concluded that quantum capacitance has nothing to say
about fractional phases. Compressibility is well defined in a correlated
liquid, and capacitive measurements have been used to extract the
chemical-potential discontinuity across a fractional gap, giving a
thermodynamic measure of the many-body gap itself---for instance at
$\nu = 1/3$ in GaAs~\cite{Eisenstein1994}. What such a measurement returns is the magnitude
of an incompressibility, not the statistics of the quasiparticles that
produce it. Fractional charge and exchange statistics require
interferometry, shot noise, or tunnelling spectroscopy, and no
thermodynamic compressibility measurement substitutes for them. The same
distinction between a density of states and an invariant that is drawn
in Section~\ref{sec:ti} for the integer case applies here, and if anything more
sharply.

\section{Conclusions}\label{sec:conclusions}

The quantum Hall effect and topological insulators represent,
respectively, the field-driven and zero-field routes to the same
underlying principle: a bulk topological invariant enforces a protected,
gapless boundary state with a characteristic density of states. Quantum
capacitance, though originally introduced as a device-physics correction
to ordinary parallel-plate capacitance in 2DEGs~\cite{Luryi1988}, is well suited
to reading out this DOS structure, because it is a near-equilibrium
electrostatic measurement rather than a transport measurement. In the
QHE, quantum capacitance oscillations confirm the discrete,
macroscopically degenerate Landau-level bulk DOS that underlies the
quantized Hall plateaus~\cite{Smith1985}. In topological insulators, quantum
capacitance measurements have directly extracted the Dirac velocity and
mapped the linear boundary DOS associated with protected surface states,
and have been used to disentangle bulk, depletion, and surface
contributions in real devices~\cite{Inhofer2018,Xiu2012}. Pushed further, local
impedance measurements inside the quantum Hall bulk itself, read in
terms of quantum capacitance, show that the textbook picture of a
spatially uniform incompressible bulk is an idealization:
self-consistent screening produces a patchwork of compressible and
incompressible regions that rearranges with magnetic field even while
the Hall plateau stays exactly quantized~\cite{Siddiki2004,Kendirlik2017}---a description
that complements, rather than replaces, the conventional
disorder--localization account of plateau formation~\cite{Aoki1981,Prange1981,Halperin1982,Huckestein1995}. The
broader lesson is that experimentally verifying a topological phase
requires probes attuned to what topology actually constrains, and
quantum capacitance, precisely because it measures compressibility
rather than conductance, has become one of the most direct such probes,
both across the field-induced and zero-field families of topological
matter and within the bulk of the quantum Hall system itself. Topology
constrains the existence and protection of the boundary states; quantum
capacitance is what probes their associated density of
states/compressibility.

\section*{Declarations}
\noindent\textbf{Funding:} This research received no external funding.

\noindent\textbf{Institutional Review Board Statement:} Not applicable.

\noindent\textbf{Informed Consent Statement:} Not applicable.

\noindent\textbf{Data Availability Statement:} No new data were created or analyzed in this study.

\noindent\textbf{Acknowledgments:} I sincerely thank my PhD supervisor, the late Rolf R. Gerhardts, for all his contributions to screening theory. During the preparation of this manuscript, the author used Claude, ChatGPT, Jenni AI and Grammarly for calculations, figure design and English-language improvement. The author has reviewed and edited the output and takes full responsibility for the content of this publication.

\noindent\textbf{Conflicts of Interest:} The author declares no conflict of interest.

\renewcommand{\refname}{References}


\begin{thebibliography}{99}\small
\bibitem{vonKlitzing1980} von Klitzing, K.; Dorda, G.; Pepper, M. New Method for High-Accuracy Determination of the Fine-Structure Constant Based on Quantized Hall Resistance. \emph{Phys. Rev. Lett.} \textbf{1980}, \emph{45}, 494--497.
\bibitem{Tsui1982} Tsui, D.C.; St\"ormer, H.L.; Gossard, A.C. Two-Dimensional Magnetotransport in the Extreme Quantum Limit. \emph{Phys. Rev. Lett.} \textbf{1982}, \emph{48}, 1559--1562.
\bibitem{Thouless1982} Thouless, D.J.; Kohmoto, M.; Nightingale, M.P.; den Nijs, M. Quantized Hall Conductance in a Two-Dimensional Periodic Potential. \emph{Phys. Rev. Lett.} \textbf{1982}, \emph{49}, 405--408.
\bibitem{Kane2005} Kane, C.L.; Mele, E.J. $Z_2$ Topological Order and the Quantum Spin Hall Effect. \emph{Phys. Rev. Lett.} \textbf{2005}, \emph{95}, 146802.
\bibitem{Bernevig2006} Bernevig, B.A.; Hughes, T.L.; Zhang, S.-C. Quantum Spin Hall Effect and Topological Phase Transition in HgTe Quantum Wells. \emph{Science} \textbf{2006}, \emph{314}, 1757--1761.
\bibitem{Koenig2007} K\"onig, M.; Wiedmann, S.; Br\"une, C.; Roth, A.; Buhmann, H.; Molenkamp, L.W.; Qi, X.-L.; Zhang, S.-C. Quantum Spin Hall Insulator State in HgTe Quantum Wells. \emph{Science} \textbf{2007}, \emph{318}, 766--770.
\bibitem{Zhang2009} Zhang, H.; Liu, C.-X.; Qi, X.-L.; Dai, X.; Fang, Z.; Zhang, S.-C. Topological Insulators in Bi$_2$Se$_3$, Bi$_2$Te$_3$ and Sb$_2$Te$_3$ with a Single Dirac Cone on the Surface. \emph{Nat. Phys.} \textbf{2009}, \emph{5}, 438--442.
\bibitem{XiaY2009} Xia, Y.; Qian, D.; Hsieh, D.; Wray, L.; Pal, A.; Lin, H.; Bansil, A.; Grauer, D.; Hor, Y.S.; Cava, R.J.; et al. Observation of a Large-Gap Topological-Insulator Class with a Single Dirac Cone on the Surface. \emph{Nat. Phys.} \textbf{2009}, \emph{5}, 398--402.
\bibitem{Hasan2010} Hasan, M.Z.; Kane, C.L. Colloquium: Topological Insulators. \emph{Rev. Mod. Phys.} \textbf{2010}, \emph{82}, 3045--3067.
\bibitem{Laughlin1981} Laughlin, R.B. Quantized Hall Conductivity in Two Dimensions. \emph{Phys. Rev. B} \textbf{1981}, \emph{23}, 5632--5633.
\bibitem{Buttiker1988} B\"uttiker, M. Absence of Backscattering in the Quantum Hall Effect in Multiprobe Conductors. \emph{Phys. Rev. B} \textbf{1988}, \emph{38}, 9375--9389.
\bibitem{Aoki1981} Aoki, H.; Ando, T. Effect of Localization on the Hall Conductivity in the Two-Dimensional System in Strong Magnetic Fields. \emph{Solid State Commun.} \textbf{1981}, \emph{38}, 1079--1082.
\bibitem{Prange1981} Prange, R.E. Quantized Hall Resistance and the Measurement of the Fine-Structure Constant. \emph{Phys. Rev. B} \textbf{1981}, \emph{23}, 4802--4805.
\bibitem{Halperin1982} Halperin, B.I. Quantized Hall Conductance, Current-Carrying Edge States, and the Existence of Extended States in a Two-Dimensional Disordered Potential. \emph{Phys. Rev. B} \textbf{1982}, \emph{25}, 2185--2190.
\bibitem{Huckestein1995} Huckestein, B. Scaling Theory of the Integer Quantum Hall Effect. \emph{Rev. Mod. Phys.} \textbf{1995}, \emph{67}, 357--396.
\bibitem{Qi2011} Qi, X.-L.; Zhang, S.-C. Topological Insulators and Superconductors. \emph{Rev. Mod. Phys.} \textbf{2011}, \emph{83}, 1057--1110.
\bibitem{Luryi1988} Luryi, S. Quantum Capacitance Devices. \emph{Appl. Phys. Lett.} \textbf{1988}, \emph{52}, 501--503.
\bibitem{XiaJ2009} Xia, J.; Chen, F.; Li, J.; Tao, N. Measurement of the Quantum Capacitance of Graphene. \emph{Nat. Nanotechnol.} \textbf{2009}, \emph{4}, 505--509.
\bibitem{Smith1985} Smith, T.P.; Goldberg, B.B.; Stiles, P.J.; Heiblum, M. Direct Measurement of the Density of States of a Two-Dimensional Electron Gas. \emph{Phys. Rev. B} \textbf{1985}, \emph{32}, 2696--2699.
\bibitem{Inhofer2018} Inhofer, A.; Duffy, J.; Boukhicha, M.; Bocquillon, E.; Palomo, J.; Watanabe, K.; Taniguchi, T.; Est\`eve, I.; Berroir, J.-M.; F\`eve, G.; Pla\c{c}ais, B.; Assaf, B.A. RF Quantum Capacitance of the Topological Insulator Bi$_2$Se$_3$ in the Bulk Depleted Regime for Field-Effect Transistors. \emph{Phys. Rev. Appl.} \textbf{2018}, \emph{9}, 024022.
\bibitem{Xiu2012} Xiu, F.; Meyer, N.; Kou, X.; He, L.; Lang, M.; Wang, Y.; Yu, X.; Fedorov, A.V.; Zou, J.; Wang, K.L. Quantum Capacitance in Topological Insulators. \emph{Sci. Rep.} \textbf{2012}, \emph{2}, 669.
\bibitem{Tahir2013} Tahir, M.; Sabeeh, K.; Schwingenschl\"ogl, U. Quantum Capacitance of an Ultrathin Topological Insulator Film in a Magnetic Field. \emph{Sci. Rep.} \textbf{2013}, \emph{3}, 1261.
\bibitem{Chklovskii1992} Chklovskii, D.B.; Shklovskii, B.I.; Glazman, L.I. Electrostatics of Edge Channels. \emph{Phys. Rev. B} \textbf{1992}, \emph{46}, 4026--4034.
\bibitem{Siddiki2004} Siddiki, A.; Gerhardts, R.R. Incompressible Strips in Dissipative Hall Bars as Origin of Quantized Hall Plateaus. \emph{Phys. Rev. B} \textbf{2004}, \emph{70}, 195335.
\bibitem{Ilani2004} Ilani, S.; Martin, J.; Teitelbaum, E.; Smet, J.H.; Mahalu, D.; Umansky, V.; Yacoby, A. The Microscopic Nature of Localization in the Quantum Hall Effect. \emph{Nature} \textbf{2004}, \emph{427}, 328--332.
\bibitem{Kendirlik2017} Kendirlik, E.M.; Sirt, S.; Kalkan, S.B.; Ofek, N.; Umansky, V.; Siddiki, A. The Local Nature of Incompressibility of Quantum Hall Effect. \emph{Nat. Commun.} \textbf{2017}, \emph{8}, 14082.
\bibitem{Rosen2022} Rosen, I.T.; Andersen, M.P.; Rodenbach, L.K.; Tai, L.; Zhang, P.; Wang, K.L.; Kastner, M.A.; Goldhaber-Gordon, D. Measured Potential Profile in a Quantum Anomalous Hall System Suggests Bulk-Dominated Current Flow. \emph{Phys. Rev. Lett.} \textbf{2022}, \emph{129}, 246602.
\bibitem{Ferguson2023} Ferguson, G.M.; Xiao, R.; Richardella, A.R.; Low, D.; Samarth, N.; Nowack, K.C. Direct Visualization of Electronic Transport in a Quantum Anomalous Hall Insulator. \emph{Nat. Mater.} \textbf{2023}, \emph{22}, 1100--1105.
\bibitem{Laughlin1983} Laughlin, R.B. Anomalous Quantum Hall Effect: An Incompressible Quantum Fluid with Fractionally Charged Excitations. \emph{Phys. Rev. Lett.} \textbf{1983}, \emph{50}, 1395--1398.
\bibitem{Jain2007} Jain, J.K. \emph{Composite Fermions}; Cambridge University Press: Cambridge, UK, 2007.
\bibitem{Jacak2022} Jacak, J.E. Formal Derivation of the Laughlin Function and Its Generalization for Other Topological Phases of FQHE. \emph{Sci. Rep.} \textbf{2022}, \emph{12}, 616.
\bibitem{Jacak2023} Jacak, J.E. Forbidden Trajectories for Path Integrals. \emph{Phys. Rev. A} \textbf{2023}, \emph{107}, 032207.
\bibitem{Eisenstein1994} Eisenstein, J.P.; Pfeiffer, L.N.; West, K.W. Compressibility of the Two-Dimensional Electron Gas: Measurements of the Zero-Field Exchange Energy and Fractional Quantum Hall Gap. \emph{Phys. Rev. B} \textbf{1994}, \emph{50}, 1760--1778.
\bibitem{Xie2021} Xie, Y.; Pierce, A.T.; Park, J.M.; Parker, D.E.; Khalaf, E.; Ledwith, P.; Cao, Y.; Lee, S.H.; Chen, S.; Forrester, P.R.; Watanabe, K.; Taniguchi, T.; Vishwanath, A.; Jarillo-Herrero, P.; Yacoby, A. Fractional Chern Insulators in Magic-Angle Twisted Bilayer Graphene. \emph{Nature} \textbf{2021}, \emph{600}, 439--443.
\end{thebibliography}
\end{document}